# Development and validation of the numerical model of Electron Cyclotron Resonance Ion Sources

V. Mironov[1], S. Bogomolov, A. Bondarchenko, A. Efremov, V. Loginov, D. Pugachev

*Joint Institute for Nuclear Research, Flerov Laboratory of Nuclear Reactions,*

*Dubna, Moscow Reg. 141980, Russia*

E-mail: vemironov@jinr.ru

Processes of the secondary electron emission (SEE) from the walls are included into the Numerical Advanced Model of Electron Cyclotron Resonance Ion Sources (NAM-ECRIS). It is found that SEE strongly influences electron confinement time and ion production. With the modified model, we observe reactions of the source to changes in a gas flow into the source and in an injected microwave power. The source performance with scaling the hexapole magnetic field is investigated. The calculated tendencies are close to the experimental observations.

## Content

---

[1] The corresponding author.

# 1. Introduction

The Electron Cyclotron Resonance Ion Sources (ECRIS) are versatile tools for the production of intense beams of highly charged ions. Plasma in ECRIS is heated by resonant absorption of microwaves and is confined in the minimum-B magnetic trap, with a plasma density of around critical density (~$10^{12}$ cm$^{-3}$), a plasma electron lifetime close to 1 ms, and electron energies in the keV range. In these conditions, moderately and highly charged ions are produced by sequential ionizing collisions of the plasma electrons with heavy particles, to be used for various applications [1].

The plasma in ECRIS is strongly influenced by the source conditions, such as the gas flow, injected microwave power, magnetic field configuration, etc. To increase the extracted ion currents and their stability, it is valuable to numerically simulate the plasma processes. The NAM-ECRIS model [2,3] is one of the advanced tools in this field. The model traces the ion and electron dynamics by accounting for various mechanisms that impact the particle densities and energies, including 3-D spatial distributions inside the ECRIS plasma. Despite the general success of the model in predicting the source behavior in the almost self-consisting way with a minimum of free parameters, some important features are still not accounted for. In particular, calculations of the electromagnetic field propagation and absorption in the plasma are done with COMSOL Multiphysics software [4] in the cold-plasma approximation. Also, till now, processes of the secondary electron emission from the walls generated by the impact of primary electrons had been neglected in the model.

Meanwhile, the important role of the secondary electron fluxes from the walls was recognized as the potential explanation for the wall-coating effect [5], that is, an increase of extracted currents of highly charge states if the source chamber walls are covered by $Al_2O_3$ oxide layer or are made of aluminum. The positive reaction of the extracted currents is observed experimentally with insertion into the source of the specially designed metal-dielectric structures with enhanced secondary electron emission coefficient [6]. The biased electrode effect is often explained in terms of the increased flux of the secondary electrons into the plasma [7] resulting in a boost in the plasma density along the source axis.

To evaluate the importance of the secondary electrons for the ECRIS performance, we include the SEE processes into the electron module of NAM-ECRIS. In our 3-D PIC model, the emission is treated by using the Furman-Pivi model [8] for the stainless-steel surface. It is found that the SEE process strongly increases the electron lifetime in the plasma and affects the ion production. The modified model is applied for studying the basic responses of ECRIS plasma to changes in the gas flow into the source and in the injected microwave power. The obtained dependencies are consistent with the experimental observations. To further validate the model, we calculate the plasma parameters by scaling the hexapole field by factors of 0.75 and 1.25 from the nominal value and confirm that the optimal field is close to $2B_{ECR}$, as predicted by the semi-empirical scaling laws [1].

The paper is structured in the following way: first, we briefly describe the basic features of our NAM-ECRIS model; then, an implementation of the Furman-Pivi algorithm is discussed. The main parameters of the plasma with the included SEE processes are described, followed by the presentation of these parameters for different gas flows and injected microwave powers. The next section deals with hexapole field scaling and its impact on the extracted ion currents.

## 2. Basic features of the NAM-ECRIS model

The model consists of two 3D Particle-in-Cell modules that separately follow in an iterative way the electron and ion dynamics in ECRIS plasma. Particles are moving in the minimum-B magnetic trap and the electric field created by the plasma potential. During the motion, particles undergo elastic and inelastic collisions, as well as collisions with the walls. Collision rates are defined by density and energy spatial distributions of electrons and ions obtained in the previous runs of the corresponding modules. Electron movement is done in the microwave electric fields, which parameters are obtained in the COMSOL Multiphysics model for the electron densities imported from the electron module. Plasma potential is calculated by solving the 3D Poisson equation on the rectangular mesh omitting the sheath potential drop close to the walls. The sheath drop from the plasma to the walls is fixed at +25 V according to the experimental data [9]. Particle motion is affected by the electric field caused by the biased electrode voltage of -250 V close to the Ø3-cm biased electrode at the injection flange of the source and the extraction voltage of -20 kV close to the Ø1-cm extraction aperture. For numerical reasons, the sheath length is taken equal to the mesh cell length such that the potential drops are calculated in the cells adjacent to the walls. The electric field in the drop is directed along the source axis or radially, depending on the wall orientation.

The magnetic field of ECRIS consists of two components – the axially symmetric solenoidal field and the multipole component. The solenoidal field is calculated by POISSON/SUPERFISH code for the geometry of the DECRIS-PM source [10]. The fields at the axis are 1.34 T at the injection and 1.1 T at the extraction sides of the source. The hexapole field is calculated by the 3D COMSOL Multiphysics magnetostatics model. The hexapole length is equal to 26.6 cm, the inner diameter is 7.6 cm, the outer diameter is 14.4 cm. The remanent magnetization inside the NdFeB permanent magnets is set to 1.25 T; calculations are done for 24 segment Halbach configuration of the hexapole. The calculated hexapole magnetic field at the source chamber walls (at a radius of 3.5 cm) is 1.1 T at the center of the source chamber, close to the measured values. Due to the edge effects, the fields on the wall at the injection and extraction sides of the source are decreasing by ~10% for the given orientation of the hexapole and the DECRIS-PM chamber. The combination of the solenoidal and multipole fields is imported into the electron/ion modules and into the COMSOL Multiphysics model of the microwave propagation in the plasma.

Calculations begin with setting an arbitrary selected seed electron density distribution in the source (typically, uniform density inside the ECR zone) and run iteratively until the converged solution is obtained. Statistical weight of numerical particles in the electron module are selected such as to ensure that all iterations in the specific run are done for a total number of electrons in the plasma fixed at certain value in the range of around 10 μC. The number of computational particles in the electron module equals to 1000, particles are traced for ~1 ms of the physical time to ensure a stable solution and good statistics. In the ion module, the total number of computational particles is 400 000 and each iteration runs for around 10 ms of the physical time. Whenever the heavy computational particle hits the walls, it is either reflected back, or is considered to be lost and is returned through the injection port as an atom. We assume the full thermal accommodation of particle energies after collisions with the walls. The losses occur through the extraction aperture or through the pumping ports at the injection flange. The loss rate defines the gas flow into the source chamber.

In the described runs, the ion dynamics is simulated for argon only, with no mixing with another gases. There are used the single ionization rates from [11], the double ionization rates from [12], and the Langevin rates [13] for the charge-changing collisions of ions with atoms. The rates are calculated by using the electron energy distributions obtained after dividing the plasma into four different regions: close to the source axis inside and outside the ECR zone, and in the plasma halo, again inside and outside of the zone. The energy distributions are fitted by a sum of two decreasing exponents, separately for the "cold" electrons with energies up to 250 eV, for the "warm" electrons with energies in the range from 250 eV to 5 keV, and for the "hot" electrons with the energies above 5 keV. Also, ion heating rates are calculated by using these energy distribution fits.

## 3. Implementation of the Furman-Pivi model of the secondary electron emission

The Furman-Pivi (FP) model of the secondary electron emission under an impact of primary electrons on a surface is based on fitting of the experimental data for stainless-steel and copper. The SEE properties for specific surface are commonly described by the secondary emission yield (the ratio between the electron current emitted from the surface $I_{emit}$ and the impacting current $I_{imp}$), $\delta(E_0, \theta) = I_{emit}/I_{imp}$, where $E_0$ is the energy of impacting electron and $\theta$ is the angle of incidence of electron in respect to normal to the surface. In the FP model, three different mechanisms are included of the secondary electron generation ($\delta = \delta_e + \delta_r + \delta_{ts}$): 1) elastically scattered electrons $\delta_e$ with energies close to energy of the primary electron; 2) "re-diffused" electrons $\delta_r$ with broad energy spectrum in the range from $E_0$ to 0; 3) true secondary electrons $\delta_{ts}$ with energies around 5 eV. For the normal impact on the surface ($\theta=0$) the elastic scattering is maximized at low electron energies with $\delta_e \sim 0.5$. The yield of the re-diffused electrons is maximized at $\sim 200$ eV and saturates at the level of $\sim 0.75$. The true secondary electron yield is maximized at $\sim 300$ eV at $\sim 1.2$ and then relatively slow decreasing with $E_0$. All yields are increasing with $\theta$. The fits are provided for the "as received" stainless-steel with no preliminary surface conditioning or special cleaning.

We apply the FP model for description of the secondary electron emission in the NAM-ECRIS electron module. Anytime when the numerical particle crosses a computational domain boundary, there are calculated the energy of the particle and the angle of its incidence on the surface. Using the FP fits, probabilities are calculated for the particle to be elastically scattered, re-diffused, generate n true-secondary electrons or absorbed. If a particle hits the extraction aperture, it is considered to be absorbed. If re-emitted, the particle velocity and angle of emission are defined according to the FP model and the electron position is set with separation of $10^{-3}$ cm from the surface. At the next time-step, re-emitted particles are accelerated by the sheath voltage into the plasma. After leaving the sheath and entering into the plasma, the electron is scattered and have lower energy compared to the primary. Probability to hit the wall in the next bounce along the magnetic field line depends on the rate of electron heating by the microwaves and on the scattering rates in electron-electron and electron-ion collisions.

Full probability to generate the true secondary electrons is calculated, including n>1. If the number of generated true secondaries exceeds 1, then the excess value ($\delta_{ts}$ -1.0) is added to the special value $N_{ts}$ to account for the missed particle generation events, the particle parameters (velocities and position) are generated and stored in a buffer array. If particle is absorbed and $N_{ts}$ exceeds 1.0, parameters of the re-emitted particle are taken from the randomly selected index of the buffer array, and $N_{ts}$ is decremented by 1.0. If the $N_{ts}$ is lower of 1.0, the particle is considered as lost and it is returned back into computational domain with energy and position randomly selected from the array of heavy particle ionization events prepared in the ion module [2]. The number of the lost electrons defines the global electron lifetime in the plasma.

The FP model provides two options for secondary electron yield calculations – probability per incident or per penetrated electron; we select the first option. In certain conditions, the number of the generated true secondary electrons is so large that the electron lifetime is formally goes to infinity. Then, no stable solution can be obtained for the given conditions of the plasma by NAM-ECRIS, which indicates the computational limitations due to using a relatively low total number of the numerical particles in the module. In most cases, however, solution is converging after ~5 iterations.

## 4. Main parameters of the plasma with basic set of the input parameters

With the SEE processes being included into the model, we perform a series of calculations for different plasma conditions. We begin with detailed description of the solution obtained with using a basic set of the input parameters: the injected microwave of 500 W (optimal value for DECRIS-PM source), the DECRIS-PM source in the default magnetic configuration, and the total electron content in the plasma of 15 μC selected such as to maximize the calculated extracted currents of the highly charged argon ions (see next section).

The energy distributions of electrons in the plasma are fitted as a sum of decaying exponents, with the decay constants and relative contents varying for the different parts of the plasma. The distribution for the electrons inside the ECR zone along the source axis is shown in Fig.1 together with the corresponding fits. The tail extends up to ~400 keV, the exponent slopes are 2.5 keV (relative contribution 30%) and 60 keV (70%). These parameters for the axial electrons outside the ECR zone are 3.1 keV (28%) and 58 keV (72%), for the halo electrons inside the zone – 4 keV (15%) and 43 keV (85%), for the halo electrons outside the zone – 4 keV (10%) and 58 keV (90%): variation in the parameters is not large in the given conditions of the source.

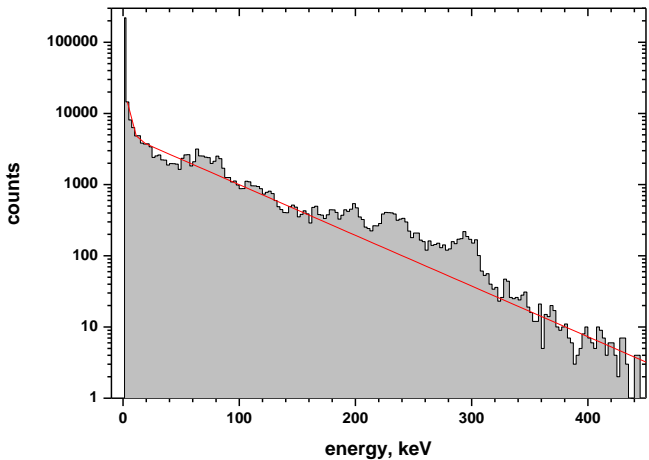

**Figure 1.** Electron energy distribution.

The low-energy part of the energy distribution is shown in Fig.2 for the axial electrons inside the ECR zone. The cold electron spectrum (up to 250 eV) is fitted with the single exponent with the slope of 275 eV, the warm electron energy distribution is a sum of exponents with the slopes of 125 eV (38%) and 1.3 keV (62%). For other parts of the plasma, the spectra are similar, with exception of the halo electrons outside the zone, where the cold electron exponential distribution decays as 28 eV, an order of magnitude smaller decay constant than in other parts. For the axial electrons inside the ECR zone, the cold electron component constitutes 42% of the total electron number, the warm component is 26%, and the hot electrons – 32%. The halo plasma electrons are noticeably hotter: there, the cold electron population outside the zone is 9% only, the warm – 13%, and the hot – 78%.

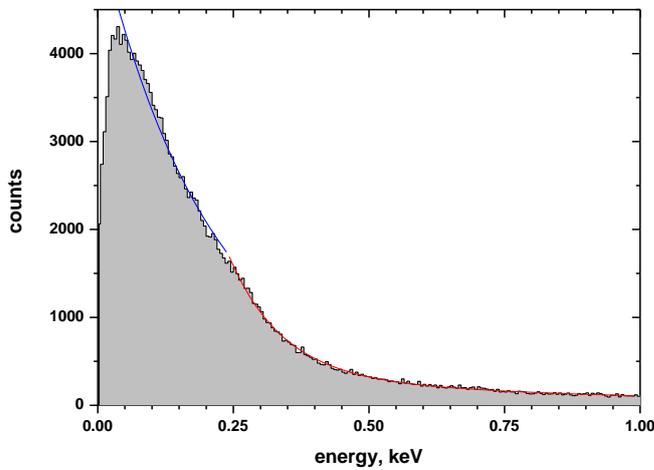

**Figure 2.** Low energy part of the electron energy distribution.

The electron density is strongly peaked on the source axis because the electrons are electrostatically trapped between the negatively biased electrode at the injection side of the source and extraction aperture at the extraction side. The electron density profile along the radial direction (along x) at the plasma chamber center is shown in Fig.3 separately for the cold (black), warm (red) and hot electrons (green), and for the total density (blue). The radial size of the central core plasma is around 1 cm, the plasma maximum density is $5.5 \times 10^{12}$ cm$^{-3}$, that is, the plasma is strongly over-dense (the critical density for the 14.5 GHz microwaves is $2.6 \times 10^{12}$ cm$^{-3}$). Mild hollowing of the warm and hot electron component distributions is observed.

The longitudinal distribution of electron density along the source axis is shown in Fig.4, again separately for the cold, warm, hot electrons and for the total density. Axial localization of electrons inside the ECR zone is less pronounced in comparison to our previous calculations [3], mostly because of the larger cold electron density in the present conditions, which results in lower heating rate of electrons close to the resonance surface.

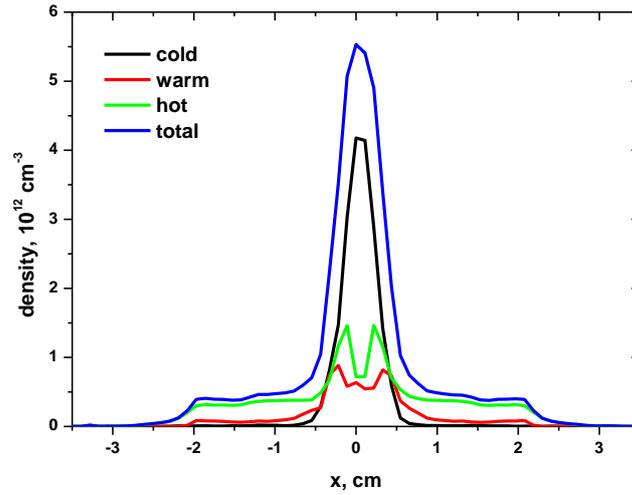

**Figure 3.** Radial distributions of the cold (black), warm (red), hot (green) electron densities and total electron density (blue).

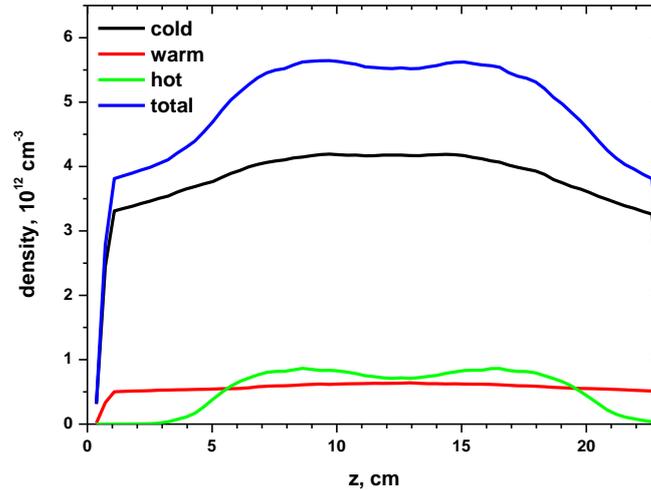

**Figure 4.** Axial distributions of the cold (black), warm (red), hot (green) electron densities and total electron density (blue).

The global electron life-time is calculated to be 0.4 ms in the basic conditions, substantially longer than it was obtained in [3]. If we calculate the plasma parameters in the same basic conditions but without including the secondary electron emission, then the electron life-time is 0.14 ms – the SEE processes increase the life-time by a factor of almost 3.

On average, 30% of the electrons that collide with the walls are elastically reflected, 40% are re-diffused from the surface, and 20% generate the true secondary electrons with an average emission of 1.25 electrons per impact. The remaining 10% of the impinging electrons are absorbed at the walls and leave the plasma; flux of these absorbed electrons out of the stationary plasma is balanced by electrons created in the ionizing collisions (~50% in the basic conditions) and by the excess (n>1) secondary electron emission (the remaining 50%).

The plasma potential is building such as to ensure the charge quasi-neutrality in the plasma. Strong localization of the ECRIS plasma along the source axis and relatively large content of cold electrons results in a decrease of the plasma potential close to the axis. The radial distribution of

– 7 –

the plasma potential is shown in Fig.5; the longitudinal distribution is shown in Fig.6. The plasma potential dip is calculated as 1.5 V, with the radial size of the potential depression of ~1 cm being defined by the plasma radial size. Longitudinally, plasma potential drops at a distance of a few mm from the walls. The relatively large gradients of the plasma potential and large cold electron content result in effective ion heating. The calculated temperatures of ions inside the plasma are increasing with the ion charge state almost linearly from 0.3 eV for $Ar^{1+}$ to 2.4 eV for $Ar^{10+}$ in the given conditions (2 eV for $Ar^{8+}$).

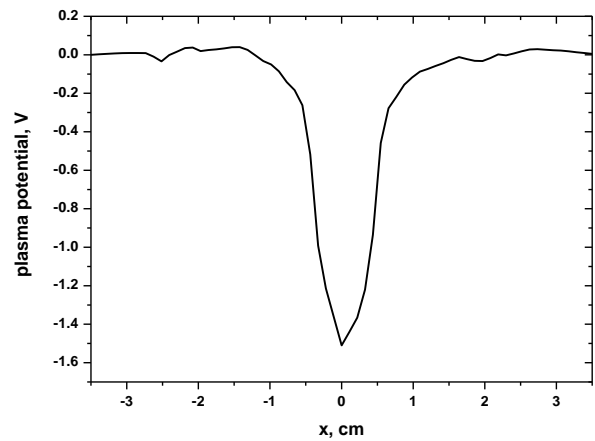

**Figure 5.** Radial distribution of the plasma potential

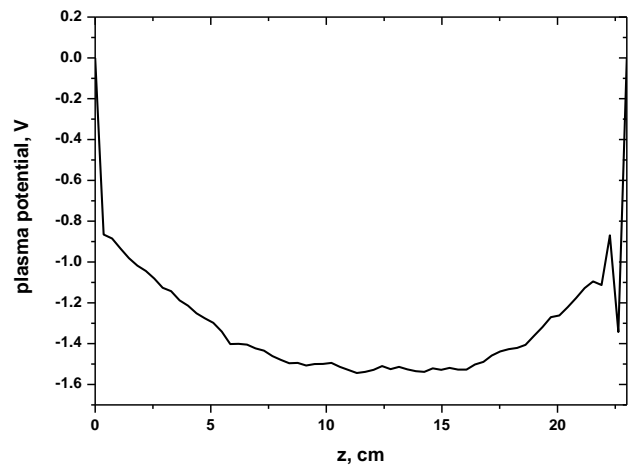

**Figure 6.** Axial distribution of the plasma potential.

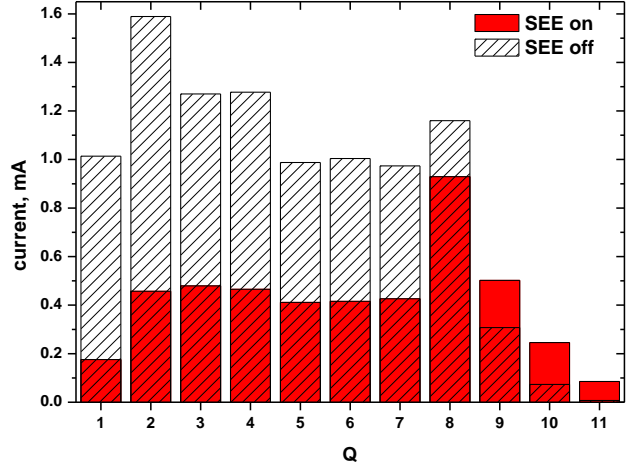

**Figure 7.** Charge state distribution of the extracted argon ions in the basic conditions (red columns) and without secondary electron emission (hatched columns).

In the ion module, the gas flow into the source should be 2.2 p-mA to make the ion lifetime equal to the electron lifetime of 0.4 ms. At that, the total extracted ion current is 4.6 mA and the particle flow into the extraction aperture is 1.1 p-mA, with the 50% of the injected particles going out of the source chamber through the pumping ports at the injection flange. The charge state distribution of the extracted argon ions is shown in Fig.7 as red columns. The distribution with the suppressed secondary electron emission is shown as the hatched columns. The gas flow for the "SEE-off" case is 8.2 p-mA, much larger than in the basic conditions. For the basic case with the enabled SEE process, the ion currents are maximized at the argon charge state (8+) at 0.9 mA. The secondary electron emission processes allow increasing the ion output for the highest charge state, above 9+ for argon. Generally, the calculated currents are close to the experimentally observed values [10]. The extraction efficiency for the $Ar^{8+}$ ions is 30%, being defined as the ratio between the extracted flux of these ions and total flux of ions out of the plasma.

## 5. Reactions of the plasma to changes in the gas flow, the injected microwave power and the hexapole field.

The gas pressure (gas flow into the source) and the injected microwave power have a strong effect on ECRIS performance [1]. Experimentally, it is observed that the low gas flows are beneficial for production of highest charge states of ions in ECRIS, with saturation or decrease of the extracted ion currents if the gas flow into the source is too high. Often, this effect is ascribed to charge-exchange processes that decrease the ion charge state in collisions of ions with neutral particles [1]. Also, it is observed in numerical simulations that the plasma potential gradients increase with the gas flow and, if too high, can cause the ion radial transport out of the plasma axis resulting in a decrease of the extracted currents [2].

Increase in the injected microwave power in most cases also increases the output of highly charged ions up to a certain level, with saturation in currents if the power is too high. It is conjectured that the electron energies and confinement are affected by the microwave power [14], thus influencing the ionization rates and the plasma density. Plasma instabilities can be excited at the high-power levels resulting in a loss of plasma confinement [15] and degradation of the source performance.

### 5.1. Gas flow

We study the effects of changing the gas flow by comparing the calculated plasma parameters with setting three different total electron contents – the basic 15 µC, low 11.7 and large 18.3 µC. The injected microwave power is fixed at 500 W. It is obtained that the gas flows into the source should be 2.2, 1.65 and 3.5 p-mA respectively for solutions to converge. The calculated charge state distributions of the extracted argon ion currents are shown in Fig.8, where the red columns are the currents for the basic conditions, the orange columns represent the case for the lowest electron content, and the white hatch-filled columns are for the case with highest electron content.

The experimentally observed tendency of decreasing currents of the highest charge states (10+ and above) with the increasing gas flow is reproduced. The lower charge state (less than 7+) currents are increasing with the gas flow. The mean charge state of the extracted ions decreases from 4.25 to 3.4 for the investigated range of the gas flow variation, while the total extracted current increases from 3.3 to 4.7 mA showing some saturation. The electron density in the center of the plasma is $4.5 \times 10^{12}$ cm$^{-3}$ for the low gas flow and it decreases to $2.2 \times 10^{12}$ cm$^{-3}$ for the high flow and large electron content. This density drop in the center is caused by the wider spatial plasma distribution, with higher densities in the plasma halo and weaker concentration of the plasma inside the ECR zone along the source axis.

The electron lifetime is largest for the low gas flow (0.47 ms) and decreases to 0.28 ms for the large flow. Variations in the hot electron energy distributions are not significant, with the tail slopes being around 40-60 keV. The cold electron distributions differ substantially: for the low gas flow, the distribution is almost constant in the range of up to 250 eV, while for the large flow the distribution is exponentially decreasing in this range with the slope of 140 eV. The relative contributions of cold/warm/hot electron components vary strongly with the gas flow: for the low flow, the cold electron contribution in the axial plasma inside the ECR zone is 10%, the warm – 35% and the hot - 55%, while for the large flow the cold contribution increases up to 55%, the warm one decreases to 20% and the hot component is 25% only. The mean energy of the combined cold and warm components inside the zone on axis is 1.4 keV for the low flux, 0.5 keV for the basic conditions and 0.4 keV for the largest gas flow. The observed cooling of low and medium energy of electron population in the plasma with the large gas flow results in less significant contribution of the secondary electron emission in the total flux of electrons into the plasma, 28% compared to 50% for the lower flows.

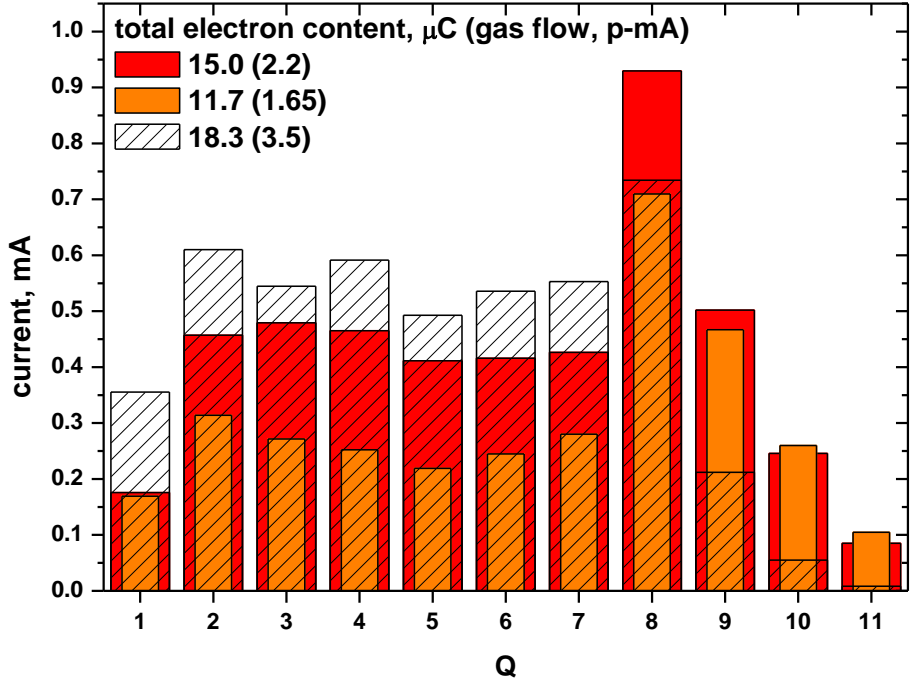

**Figure 8.** Charge state distributions of the extracted argon ions for the basic conditions (red columns, gas flow of 2.2 p-mA), for the low gas flow (orange columns, 1.65 p-mA) and for the large gas flow (hatched columns, 3.5 p-mA).

Wider spatial distribution of the plasma density for the large gas flow decreases the extraction efficiency of highly charged ions: for $Ar^{8+}$ ions, the efficiency is 18%. At that, plasma potential dips are close each other for the cases with the low and large gas flows, ~ 0.5 V. Ion temperatures for the $Ar^{8+}$ ions are ~0.5 eV for both cases, less than in the basic conditions.

### 5.2. Microwave power

The charge state distributions for different levels of the injected microwave power are shown in Fig.9. The orange columns are for the case with 250 W of the power for the total electron content of 15 μC, same as in the basic conditions (the red columns). The hatched columns represent the currents for 1000 W of the injected power and the total electron content of 21.7 μC, much larger than the basic one. The reason for the increased electron content is our intention to compare the extracted currents for approximately the same gas flow into the system. For the electron content of 15 μC and 1 kW of the power, the electron lifetime is ~0.85 ms, the plasma density in the center is $5\times10^{12}$ $cm^{-3}$, and the gas flow into the source is as low as 0.75 p-mA, a factor of ~3 lower than in the basic case. For the larger electron content, the lifetime is decreasing down to 0.43 ms, electron density remains ~ $5\times10^{12}$ $cm^{-3}$ and the gas flow is 2.4 p-mA, close to the basic value.

For the low injected power, the calculated gas flow is 2.5 p-mA and there is no need to adjust the electron content to obtain solution with the same level of the flow. The electron density at the center of the plasma is $6\times10^{12}$ $cm^{-3}$, that is, this value is not changing substantially with the microwave power. The electron lifetime for the low power is decreasing to 0.36 ms, and the cold electron energy distribution is decreasing with the tail slope of 50 eV, much faster than in the case of the large microwave power (240 eV). Also, the hot electron distribution is less energetic at the low microwave power: the slope of the distribution is 35 keV compared to 80 keV for the 1-kW case. Contributions of different electron components into the total electron population on axis inside the ECR zone are 75%(cold)+8%(warm)+17%(hot) for the 250-W case, and 22%(cold)+14%(warm)+64%(hot) for 1 kW of the power.

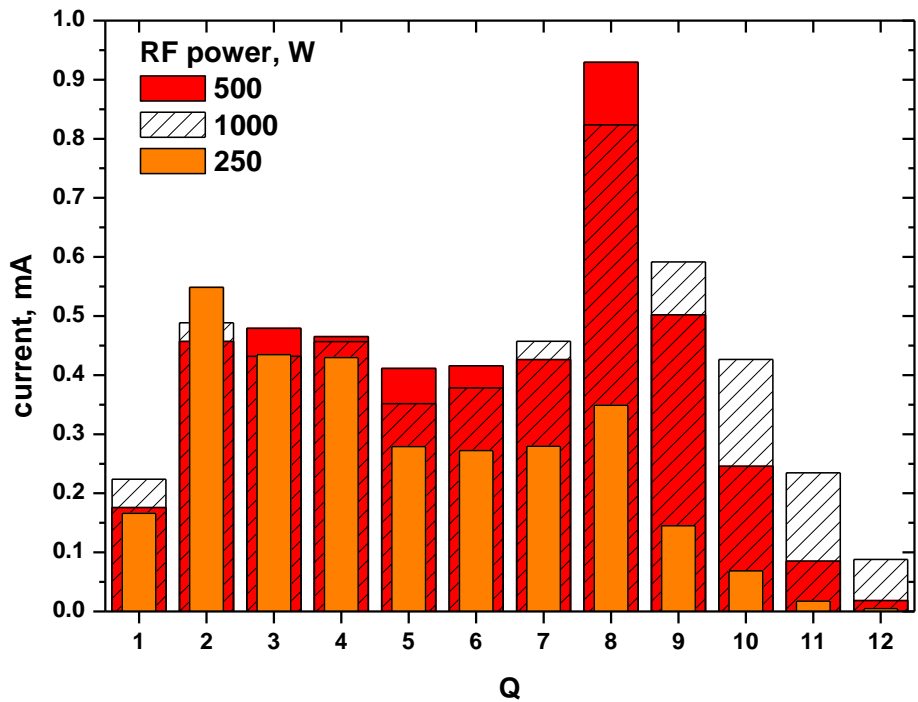

**Figure 9.** Charge state distributions of the extracted argon ions for the basic conditions (red columns, the injected microwave power of 500 W), for the low injected microwave power (orange columns, 250 W) and for the large injected power (hatched columns, 1000 W).

These variations in the electron energy distributions with the microwave power affect the extracted ion currents. For the low power, extracted currents for the highest argon charge states are smaller by a factor of ~5 compared to the default 500 W case. Increase in power up to 1 kW shifts the charge state distribution to the higher charge states in correspondence to the experimental observations.

Power balance of electrons is mostly defined by two counter-acting processes, electron heating by the microwaves and energy losses in collisions with the walls. To estimate how fast are these processes, we calculate the time dependence of the electron mean energy for the cold and warm populations after termination of the microwave heating. The calculations are done for 1 kW of the microwave power and the electron content of 15 μC. The dependence is shown in Fig.10 (black curve, left scale), where we also plot the time dependencies of the number of events of the electron collisions with the walls (blue, left scale) and of the electron losses from the plasma (red, left scale, the value is multiplied by factor of 10 to fit into the plot). The moment of terminating the microwave heating is shown by the vertical line. Immediately after RF-heating termination, the electron mean energy starts to decrease and reaches the level of 200 eV (50% of the initial value) in ~100 μs.

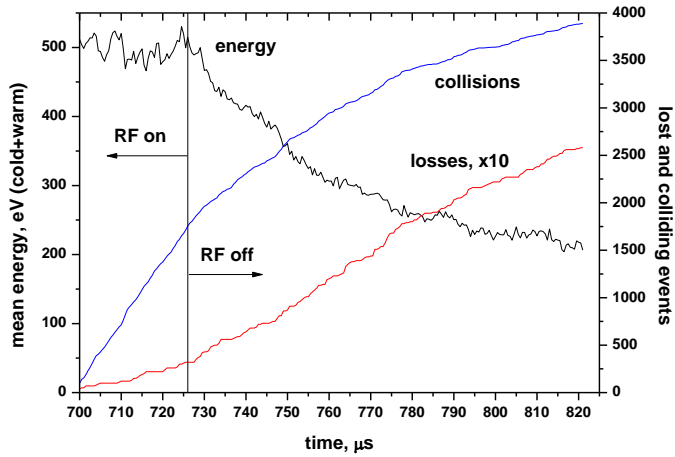

**Figure 10.** Mean energy of the combined cold and warm electron populations (black, left scale), total number of wall-collision (blue, right scale) and wall-loss events (red, right scale, multiplied by 10) as functions of time before and after switching the microwave heating off.

Simultaneously, frequency of electron collisions with the walls (defined by a slope of the blue curve) is decreasing by a factor of ~2 due to termination of the RF-induced velocity diffusion of electrons and smaller flux of energetic electrons out of the plasma (see also measurements reported in [16]). However, this decrease does not necessarily result in a decrease of the electron losses. Instead, in the presented conditions the loss frequency is increasing, mostly because of suppressed emission of the secondary electrons generated in collisions of the energetic electrons with the walls. Before switching the RF-heating off, the secondary electrons contribute 70% into the total flux of electrons into the plasma in these conditions, and this contribution drops to 35% in 100 μs after termination of the heating, which results is the decreased electron lifetime.

The NAM-ECRIS model is stationary and can not be applied for studies of the transitional processes in ECRIS plasma. Nonetheless, the calculated tendencies of the fast electron cooling after RF-heating termination are useful for analysis of the afterglow effect – the experimentally observed fast increase in the extracted currents of the highly charged ions soon after switching the microwave injection off [17]. The electron cooling implies the increased rates of ion heating in electron-ion collisions, which can result in faster escape of highly charged ions out of the plasma.

### 5.3. Hexapole field scaling

Further validation of the model is done by comparing the source output for different hexapole fields. According to the semi-empirical scaling laws [1], the extracted currents are steadily increasing with the hexapole fields until reaching some saturation. Increasing the field above the optimal value either do not affect the currents, or even results in their decrease. To check this tendency, we calculate the source plasma parameters for different hexapole fields with the same total electron content in the plasma of 15 μC and with the fixed microwave power of 500 W. The results for the charge-state-distributions of the extracted ion currents are shown in Fig.11 for the basic conditions (the red columns), for the hexapole field scaled with the factor of 0.75 (the orange columns, field at the source chamber walls is 0.825 T) and for the field scaling of 1.25 (the hatched columns, the field is 1.375 T). The decrease of the hexapole field results in a pronounced drop in the extracted currents of the highly charged ions, the increased field moderately enhances the currents for the highest charge states, with almost no gain close to the maximum of the distribution at $Ar^{8+}$.

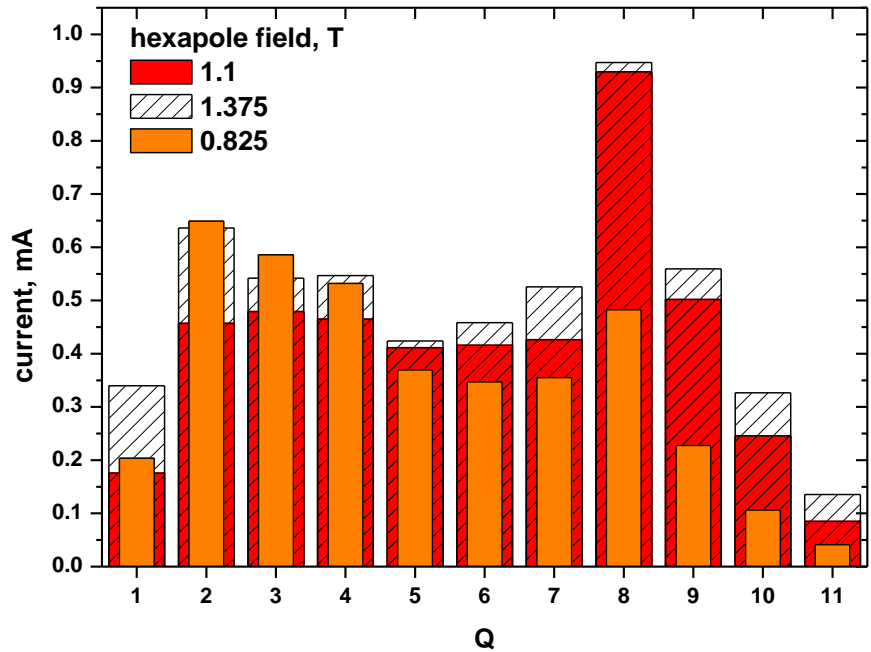

**Figure 11.** Charge state distributions of the extracted argon ions for the basic conditions (red columns, the hexapole field of 1.1 T), for the low injected microwave power (orange columns, 0.825 T) and for the large injected power (hatched columns, 1.375 T).

For the low hexapole field, the electron lifetime is 0.42 ms, the electron density at the plasma center is $5.5 \times 10^{12}$ cm$^{-3}$, the hot electron energy distribution tail slope is 44 keV, and the gas flow is 2.5 p-mA. These values are close to the values obtained in the basic conditions. The major changes are seen in the electron energies of the cold electron component: the energy distribution slope is 110 eV for the axial plasma inside the ECR zone, substantially lower than in the basic conditions of the plasma. The relative contributions of different components are 75% (cold)+10% (warm)+15% (hot), which makes the electron population much colder compared to the basic one. The results are that the plasma potential is 2.5 V at the axis, the ions are strongly heated (the temperature for $Ar^{8+}$ ions is around 3.3 eV) and extraction efficiency for $Ar^{8+}$ ions is 24%, indicating less effective transport of ions from the dense parts of the plasma into the extraction aperture. Combined with the lower ionization rates for the highly charged ions, this results in the general reduction of the extracted ion currents.

For the large hexapole field, the electron lifetime is the relatively small 0.3 ms. The electron density is not changed ($5.0 \times 10^{12}$ cm$^{-3}$ at the center), as well as the electron energy distributions with the slopes of 66 keV for the hot electrons and 180 eV for the cold ones. The relative contributions of the components in the central plasma are 30% (cold)+15% (warm)+55% (hot), which indicates more energetic electron population for the larger hexapole fields. The gas flow should be increased up to 3.3 p-mA to obtain the plasma with the total content of 15 μC, the plasma potential dip is 1.0 V and the ion temperatures are relatively small (1.3 eV for $Ar^{8+}$). Extraction efficiency is not changed noticeably, 28% for $Ar^{8+}$.

The observed production of more energetic electrons for the larger hexapole field is followed by increase in the mean energies of the lost electrons: for the hexapole field of 0.825 T, the mean energy is 0.85 keV, increasing up to 3.4 keV for the default hexapole field and up to 7.7 keV for the field of 1.375 T.

It is useful to compare rates of the cold electron heating by microwaves for different hexapole fields. The calculated rates are 65, 85 and 110 eV/µs for the hexapole fields of 0.825, 1.1 and 1.375 T respectively. The larger is the hexapole field, the more effective is the electron heating, mostly due to smaller dimensions of the ECR volume in the radial direction and increased frequency of crossing the ECR surface for trapped electrons.

## 6. Discussion and conclusions

The calculated reactions of the plasma are consistent with the experimental observations. We conclude that, in the present form, the NAM-ECRIS model correctly describes the main features of the ECRIS plasma, thus allowing predicting the source performance and paving the ways for the source optimization.

The important factor that influences the plasma dynamics are parameters of the cold electron component, with strong variations in these electrons' energy distribution and relative content depending on the source conditions. In the Table 1, we compile some parameters for different source inputs, showing the slopes of energy distributions of the cold and hot electrons, the relative contributions of the components and the extracted currents of the highly charged $Ar^{10+}$ ions.

TABLE 1. MAIN PARAMETERS OF THE COLD AND HOT ELECTRON COMPONENTS FOR THE INVESTIGATED SOURCE CONDITIONS.

|  | Cold, eV | Cold, % | Hot, keV | Hot, % | $Ar^{10+}$ current, µA |
|---|---|---|---|---|---|
| Basic (SEE on, 500 W, 2.2 p-mA) | 275 | 26 | 60 | 32 | 245 |
| SEE off (500 W, 8.2 p-mA) | 76 | 41 | 43 | 42 | 73.5 |
| Low gas flow (500 W, 1.65 p-mA) | const | 10 | 45 | 55 | 260 |
| High gas flow (500 W, 3.5 p-mA) | 143 | 55 | 33 | 25 | 55 |
| Low power (250 W, 2.5 p-mA) | 50 | 75 | 33 | 17 | 69 |
| High power (1000 W, 2.4 p-mA) | 240 | 22 | 80 | 64 | 425 |
| Low hexapole (500 W, 2.4 p-mA) | 110 | 75 | 44 | 15 | 106 |
| High hexapole (500 W, 3.3 p-mA) | 180 | 30 | 66 | 55 | 326 |

It is seen that the small content of the cold electrons and their relatively high energies correlate with the large currents of the highly charged ions. The hot electron energies are changing in the range (30-80 keV) with the gas flow, the microwave power and the hexapole field. The cold electron component's dynamics is governed by the electrostatic confinement of electrons in between the biased electrode and the extraction aperture, by frequent collisions with the walls followed by re-emission of the scattered and secondary electrons, and by energization of electrons when they cross the ECR zone surface. Due to the energy dependence of the secondary electron emission yield, the higher is energy of the cold electron population, the more secondary electrons are injected into the plasma after electron losses on the source walls, and therefore the larger is the electron lifetime in the plasma. When energized, the electrons in this energy range have larger ionization rates and lower ion heating rates in electron-ion collisions, which benefits the highly charged ion production. The electron heating rates increase due to higher injected microwave power; for the low plasma density, microwave absorption in the outer parts of the plasma beyond the ECR zone is relatively small, which results in larger electric fields of the microwaves at the ECR surface and larger heating rates of electrons. These considerations explain the observed tendencies of producing more highly charged ions with the high injected microwave power and low gas flows into the source.

It was demonstrated in [3] that for the decreased solenoidal field at the source center (the decreased $B_{min}$ value), the electron confinement time in the plasma is decreasing. The effect was attributed to the transport of electrons to the halo from the central core plasma in the case when the magnetic field lines become to be more curved toward the radial walls of the source. This transport results in the boosted density of the hot electrons in the plasma and in the loss of the electrostatic confinement of electrons in the core plasma. In the halo, energetic electrons are well confined due to their low collisionality until they reach the energies of around 100 keV; above these energies the electrons are rapidly lost due to the non-adiabatic effects, which strength is proportional to the ratio between the electron Larmor radius and the magnetic field line curvature [3].

For the fixed solenoidal and increased hexapole fields, the analogous effect can be observed. Indeed, we obtain that for the hexapole field of 1.375 T the electron lifetime is much shorter compared to the basic conditions, which, in combination with the larger content of the hot electrons with their relatively small ionization rates, causes the source performance saturation for the large hexapole fields. For the low hexapole field, the extracted ion currents decrease because of the lower heating rates of electrons, increased content of the cold electrons and less effective transport of ions toward the extraction aperture.

**Acknowledgements**

This work was supported by the Russian Foundation for Basic Research under grant No. 20-52-53026/20.
[1] R. Geller, "Electron Cyclotron Resonance Ion Sources and ECR Plasma", (Institute of Physics, Bristol), 1996
[2] V. Mironov, S. Bogomolov, A. Bondarchenko, A. Efremov, V. Loginov and D. Pugachev, "Three-dimensional modelling of processes in Electron Cyclotron Resonance Ion Source", JINST **15**, P10030 (2020); https://doi.org/10.1088/1748-0221/15/10/P10030
[3] V. Mironov, S. Bogomolov, A. Bondarchenko, A. Efremov, V. Loginov and D. Pugachev, "Numerical investigations of the minimum-B effect in Electron Cyclotron Resonance Ion Source", JINST **16** P04009 (2021); https://doi.org/10.1088/1748-0221/16/04/P04009
[4] COMSOL, Inc., COMSOL Multiphysics®, version 5.4 (2020) [https://www.comsol.com]
[5] A. G. Drentje, "Techniques and mechanisms applied in electron cyclotron resonance sources for highly charged ions", Rev. Sci. Instrum. **74**, 2631 (2003); https://doi.org/10.1063/1.1569408

[6] L. Schachter, K. E. Stiebing and S. Dobrescu, "Enhanced confinement in electron cyclotron resonance ion source plasma", Rev. Sci. Instrum. **81**, 02A330 (2010); https://doi.org/10.1063/1.3267315

[7] S. Gammino, J. Sijbring, and A. G. Drentje, "Experiment with a biased disk at the K.V.I. ECRIS", Rev. Sci. Instrum. **63**, 2872 (1992); https://doi.org/10.1063/1.1142782

[8] M. A. Furman and M. T. F. Pivi, "Probabilistic model for the simulation of secondary electron emission", Phys. Rev. ST Accel. Beams **5**, 124404 (2002); https://doi.org/10.1103/PhysRevSTAB.5.124404

[9] O. Tarvainen, P. Suominen, T. Ropponen, T. Kalvas, P. Heikkinen, and H. Koivisto, "Effect of the gas mixing technique on the plasma potential and emittance of the JYFL 14 GHz electron cyclotron resonance ion source", Rev. Sci. Instrum. **76**, 093304 (2005); https://doi.org/10.1063/1.2038647

[10] S. L. Bogomolov, A. E. Bondarchenko, A. A. Efremov, K. I. Kuzmenkov, A. N. Lebedev, V. E. Mironov, V. N. Loginov, N. Yu. Yazvitsky, N. N. Konev, "Production of High-Intensity Ion Beams from the DECRIS-PM-14 ECR Ion Source", Phys. Part. Nuclei Lett., 15, 878 (2018); https://doi.org/10.1134/S1547477118070191

[11] M. Mattioli, G. Mazzitelli, M. Finkenthal, P. Mazzotta, K.B. Fournier, J. Kaastra and M.E. Puiatti." Updating of ionization data for ionization balance evaluations of atoms and ions for the elements hydrogen to germanium", J. Phys. B: At. Mol. Opt. Phys. 40 3569 (2007); https://doi.org/10.1088/0953-4075/40/18/002

[12] M. Hahn, A. Müller, and D. W. Savin, "Electron-impact Multiple-ionization Cross Sections for Atoms and Ions of Helium through Zinc", ApJ **850** 122 (2017); https://doi.org/10.3847/1538-4357/aa9276

[13] A. Dalgarno and D. R. Bates, "The mobilities of ions in their parent gases", Philos. Trans. R. Soc. London, Ser. A, **250**, 426 (1958); https://doi.org/10.1098/rsta.1958.0003

[14] S. Gammino, G. Ciavola, L. G. Celona, D. Mascali and F. Maimone, "Numerical Simulations of the ECR Heating With Waves of Different Frequency in Electron Cyclotron Resonance Ion Sources", IEEE Transactions on Plasma Science **36**, 1552 (2008); https://dx.doi.org/10.1109/TPS.2008.927288.

[15] O. Tarvainen, T. Kalvas, H. Koivisto, J. Komppula, R. Kronholm, J. Laulainen, I. Izotov, D. Mansfeld, V. Skalyga, V. Toivanen, and G. Machicoane, "Limitation of the ECRIS performance by kinetic plasma instabilities (invited)", Rev. Sci. Instrum. **87**, 02A703 (2016); https://doi.org/10.1063/1.4931716

[16] M. Sakildien, O. Tarvainen, R. Kronholm, I. Izotov, V. Skalyga, T. Kalvas, P. Jones, and H. Koivisto, "Experimental evidence on microwave induced electron losses from ECRIS plasma", Physics of Plasmas 25, 062502 (2018); https://doi.org/10.1063/1.5029443

[17] K. Langbein, Rev. Sci. Instrum. 67, 1334 (1996); http://dx.doi.org/10.1063/1.1146711